\documentclass[10pt,letterpaper]{article}
\usepackage{opex3}
\usepackage{psfrag}
\definecolor{gris}{gray}{0.3}

\definecolor{vert}{cmyk}{0.7,0,.7,0.5}
\definecolor{rouge}{cmyk}{0,1,1,.3}
\newcommand{\afterreview}[1]{#1}

\begin{document}
\title{Continuous wave photon pair generation in silicon-on-insulator waveguides and ring resonators}
\author{S. Clemmen$^{1*}$, K. Phan Huy$^2$, W. Bogaerts$^3$, R. G. Baets$^3$, Ph. Emplit$^4$ and S. Massar$^1$}
\address{$^1$ Laboratoire d'Information Quantique, CP 225,
 Universit\'{e} Libre de Bruxelles (U.L.B.),
 Boulevard du Triomphe, B-1050 Bruxelles, Belgium}
\address{$^2$  D\'{e}partement d'Optique P.M. Duffieux, Institut
FEMTO-ST, Centre National de la Recherche Scientifique UMR 6174,
Universit\'{e} de Franche-Comt\'{e}, 25030 Besan\c{c}on, France}
\address{$^3$ Department of Information Technology (INTEC), Ghent University - IMEC,  Sint-Pietersnieuwstraat 41, 9000 Gent, Belgium}
\address{$^4$ Service OPERA-Photonique, CP 194/5, Universit\'e Libre de Bruxelles (U.L.B.), avenue F.D. Roosevelt 50,  1050 Brussels, Belgium}
\vspace{0.1cm}
\begin{center}\footnotesize{\textit{* Corresponding author \url{sclemmen@ulb.ac.be}}}\end{center}
\vspace{-0.6cm}
\begin{abstract}%
\hspace{-0.3cm} Silicon waveguides are promising $\chi^3$-based photon pair sources. Demonstrations so far have been based on picosecond pulsed lasers. Here, we present the first investigation of photon pair generation in silicon waveguides in a continuous regime. The source is characterized by coincidence measurements. \afterreview{We uncover the presence of unexpected noise which had not been noticed in earlier experiments.}
Subsequently, we present advances towards integration of the photon pair source with other components on the chip. This is demonstrated by photon pair generation in a Sagnac loop interferometer and inside a micro-ring cavity. Comparison with the straight waveguide shows that these are promising avenues for improving the source. In particular photon pair generation in the micro-ring cavity yields a source with a spectral width of approximately 150~pm resulting in a spectral brightness increased by more than 2 orders of magnitude.
\end{abstract}
\ocis{(270.0270) Quantum optics; (190.4390) Nonlinear optics, integrated optics; (190.4380) Nonlinear optics, four-wave mixing}

\section{Introduction}
Silicon-on-Insulator (SOI) is an attractive \afterreview{platform} for integrated optics. The high index of silicon  enables large density integration of optical circuits which can be fabricated with CMOS-compatible technology~\cite{reviewBaets}. The strong light confinement allows non linear effects to be realized at low pump power~\cite{reviewlipson,SOIreview}, such as Raman laser~\cite{RamanLaser}, Frequency conversion~\cite{FC,lipsonFWMrings}, Intensity Modulation~\cite{IM}. Recently photon pair production based on four wave mixing was demonstrated in SOI waveguides~\cite{kumar,ntt,takesue2,harada}.
\\
\indent
\afterreview{The later experiments follow earlier work in which $\chi^3$ non linearity was used to generate photon pairs in conventional optical fibers~\cite{fiberpairs1,fiberpairs2,fiberpairs3} and photonic crystal fibers~\cite{fiberpairs4,fiberpairs5}. 
In fiber experiments Raman scattering is often an important source of noise~\cite{fiberpairs2} whereas it should be negligible in SOI waveguides (see however below) as Raman gain is restricted to a narrow bandwidth far detuned ($15.6$~THz.) from the pump.
\\
\indent
Experiments~\cite{kumar,ntt,takesue2,harada} used pulsed lasers which limits their applicability outside the research laboratory. Here we report that photon pairs can be readily generated in SOI waveguides in a Continuous Wave (CW) regime as predicted theoretically in~\cite{agrawalpairSOI}.}
\\
\indent
An appealing feature of Silicon-on-Insulator photonics is the integration possibility of various components. Up to now, SOI photon pairs sources could not be followed by long interferometers on the chip because pairs would be generated all along the circuit. For this reason, a major advance would be to localize the photon pair generation on a restricted part of the chip. We have investigated 2 different techniques to achieve this goal. The first one is to suppress the pump beam by classical interference. We used a integrated Sagnac Loop Interferometer (SLI), i.e. a specific example of a balanced interferometer, and generated the photon pairs in the interferometer. In such a balanced interferometer, the pump beam is reflected while photon pairs are randomly reflected or transmitted but they always stick together by pairs. \afterreview{This is similar to photon bunching observed in Hong-Ou-Mandel experiments~\cite{Quantumeffectinsagnac}.} At the transmission port of the interferometer, the pump power is strongly reduced, but half of the generated pair flux remains. Here we report photon pair generation in a SLI and observe a reduction of the residual pump power by $6-16$~dB compared to the straight waveguide.
\\
\indent
A second way to localize the pair generation is to use a structure that strongly enhance the pair generation so that pair generation by the residual pump power outside of that structure can be neglected. The structure we investigated is a micro-ring resonator. In such a structure, the light is accumulated which increases dramatically nonlinear effects such as  thermal and carriers bistability~\cite{bistab:priem,bistab:XuLipson,bistab:AlmeidaLipson}, frequency conversion~\cite{lipsonFWMrings}, or the efficiency of photon pair generation. Furthermore it concentrates the generated pairs on a restricted spectral bandwidth which allows for huge spectral brightness.
Here we report photon pairs in a micro-ring resonator, and observe a spectral density of pairs 2 orders of magnitude higher than in the straight waveguide. We believe this is the first observation of photon pair generation using a $\chi^3$ nonlinearity in a resonant structure.
\\
\indent
The remainder of the paper is organized as follows. First we discuss the basic theory behind our work. Then we describe our 3 different structures and the experiment. We present our results separately for the straight waveguide, the SLI and the ring resonator. We compare experimental results with theoretical predictions and we point out differences between our 3 photon pairs sources.
\section{Theoretical aspects}
\begin{figure}[!b]\begin{center}
\psfrag{x0}[cc][cc][1][0]{\footnotesize{1}}
\psfrag{x1}[cc][cc][1][0]{\footnotesize{10}}
\psfrag{x2}[cc][cc][1][0]{\footnotesize{100}}
\psfrag{x3}[cc][cc][1][0]{\footnotesize{1000}}
\psfrag{y0}[cr][cr][1][0]{\footnotesize{$10^{-3}$}}
\psfrag{y1}[cr][cr][1][0]{\footnotesize{$10^{-2}$}}
\psfrag{y2}[cr][cr][1][0]{\footnotesize{$10^{-1}$}}
\psfrag{y3}[cr][cr][1][0]{\footnotesize{$10^0$}}
\psfrag{y4}[cr][cr][1][0]{\footnotesize{$10^1$}}
\psfrag{y5}[cr][cr][1][0]{\footnotesize{$10^2$}}
\includegraphics[width=5.5cm]{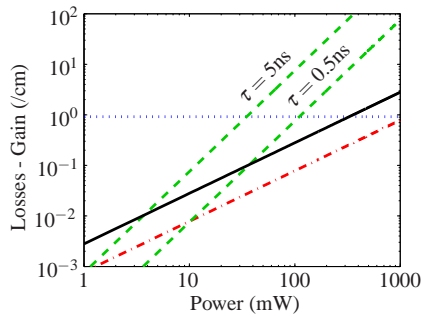}
\put(-95,-3){\footnotesize{Power (mW)}}
\put(-163,27){\rotatebox{90}{\footnotesize{Losses - Gain (/cm)}}}
\put(-60,73){\rotatebox{45}{\footnotesize{$\tau = 0.5$ns}}}
\put(-78,75){\rotatebox{45}{\footnotesize{$ \tau = 5$ns}}}
\caption{Losses due to linear scattering (\textcolor{blue}{$\cdot \cdot$}), two-photon absorption (\textcolor{red}{$\cdot -$}), and free-carrier absorption (\textcolor{green}{$--$}) for free-carrier lifetimes $\tau$ ranging from 5 to 0.5~ns. Modulational instability  gain ($\gamma P$) when phase matching is perfectly satisfied is plotted in black. In the present experiment we operate at a pump power less than $10$~mW when linear losses dominate.
}
\label{fig:loss}
\end{center}\end{figure}
In Kerr media, photon pair generation occurs through a process in which two photons from a pump beam at frequency $\omega_0$ are converted into Stokes and anti-Stokes photons at frequencies $\omega_0\pm \omega$. This can also be thought as the spontaneous amplification of vacuum fluctuations at Stokes and anti-Stokes frequencies through modulation instability. Nevertheless, it has been shown theoretically that photon pair generation differs from modulation instability by the absence of any threshold (i.e. gain does not need to be higher than losses) and even the absence of any phase matching~\cite{eddy}. For small pump power $P$ the photon pair flux $f$ (the number of photons generated per unit frequency and per unit time) at pulsation detuning $\omega$  generated over a propagation distance $L$ is given by~\cite{eddy}:
\begin{eqnarray}
f &=& \left| \frac{\gamma P}{g(\omega)}\sinh{[g(\omega)z]} \right|^2 \label{eq:flux1}\\
&\approx& (\gamma P L)^2  {\rm sinc}^2 \left(\frac{ \beta_2 \omega^2 L}{2}\right) \label{eq:flux}
\end{eqnarray}
with $\beta_2$ the group velocity dispersion of the waveguide, $g(\omega)$ a c-number representing the modulation instability "gain", $\gamma=\frac{2\pi n_2}{\lambda A_{eff}}$ the effective nonlinearity for an optical mode whose effective section is $A_{eff}$ at a wavelength $\lambda$ in a silicon waveguide with intrinsic nonlinearity $n_2$.
In Eq. (\ref{eq:flux}) we have given the expression valid to second order in $\gamma P$.
In this limit the bandwidth $\Delta \omega = \sqrt{2 \pi / |\beta_2| L}$ is independent of the pump power and of the sign of the dispersion, i.e. phase matching is not necessary, contrary to the case of modulation instability~\cite{agrawalBook}.
\\
\indent
In SOI waveguides free carrier absorption becomes important when the average power is high, which seems a justification for using  a pulsed pump as in~\cite{kumar,ntt,takesue2,harada}. Figure \ref{fig:loss} shows the evolution of nonlinear losses and Kerr nonlinearity as a function of the pump power. High power enables strong Kerr nonlinearity resulting in good efficiency of the photon pair generation process but also induces high nonlinear losses due to Free Carrier Absorption. This is the reason why pump pulses much shorter than the free carrier lifetime (which ranges from 5 to $0.5$~ns~\cite{bistab:priem}) were used in earlier experiments.  At low power, efficiency of photon pair generation is low but nonlinear losses can be neglected with respect to linear scattering loss. In the present experiment we operate at a pump power less than $10$~mW whereupon linear losses dominate.
\\
\indent
The photon pair spectral density flux generated in the ring resonator can be similarly obtained (neglecting all effects except the Kerr nonlinearity and linear losses) :
\begin{equation}
f=
( F_{p}^2 F_s F_i \gamma P L_{ring})^2
\label{eq:fluxring}
\end{equation}
In this expression $F_{p,s,i}$ are the field enhancements at the pump, signal and idler frequencies:
$F\simeq \frac{T}{ (T/2 + \eta/2)^2 + 4 \sin^2 \Phi/2}$ with $T\ll 1$ the intensity transmission coefficient of the coupler, $\eta\ll 1$ the linear intensity losses within the ring, and $\Phi$ the phase accumulated by the field during one round trip.
To derive Eq. (\ref{eq:fluxring}) we refer to~\cite{FMWinRING} where 4-wave mixing in resonant $\chi^{(3)}$ structures is studied. From the classical result one can deduce in a standard way the spontaneous effect Eq. (\ref{eq:fluxring}).
\section{Nanostructures : design and properties}
A rough description of the geometry of our 3 structures is presented as an inset in Fig.~\ref{fig:setup}.  Our structures were fabricated by the ePIXfab (\url{http://www.epixfab.eu/}) at IMEC with 193~nm deep UV lithography, so we refer to~\cite{reviewBaets,jjapbaets} and references therein for a more detailed presentation of similar structures.
\\
\indent
Waveguides have been designed to maximize photon pair generation. The design is a trade off between strong nonlinearity i.e. long propagation distance and small cross section, and low scattering loss, i.e. small propagation distance and large cross section.
\afterreview{
\subsubsection*{Straight Waveguide}
}
Our first structure is a 11.3~mm long straight waveguide and it has a section of $500\times220 \: \textrm{nm}^2$. This narrow cross section will be used everywhere we want to generate photon pairs.  Light in/output coupling with the straight waveguide is ensured by tapered sections and grating couplers. In this narrow waveguide propagation loss is $4\pm1\textrm{dB}/\textrm{cm}$; computed (commercial software FIMMWave simulation) dispersion parameter is $\beta_2=-0.7\textrm{ ps}^2/\textrm{m}$; computed effective area is $A_{eff}=0.064 \: \mu\textrm{m}^2$, resulting in $\gamma=280\:\textrm{W}^{-1}\textrm{m}^{-1}$.
\afterreview{
\subsubsection*{Sagnac Loop Interferometer}
}
The second structure we investigated is a Sagnac Loop Interferometer. It is made of a straight waveguide (10~mm long and section of $500\times220 \: \textrm{nm}^2$) linked on both end to a 3-dB beam splitter (directional coupler). This Sagnac Loop Interferometer is linked to the in/output grating couplers by 2 broad (800~nm) 2.4~mm long waveguides. Broad waveguides are chosen to decrease both nonlinearity and loss in those arms.
The directional coupler used in the Sagnac loop interferometer has been characterized thanks to transmission measurements on an isolated sample on the same silicon chip. It shows a coupling ratio estimated to be $62/38$ (rather than $50/50$) at the pump wavelength (1540~nm). Note that the coupling factor changes with wavelength to reach 52$\%$ at 1520~nm and 72$\%$ at 1560~nm. Such a coupler should induce an extinction ratio of 12~dB after the Sagnac interferometer at pump wavelength. This value is close to the measured value of $12.5\pm2.5$~dB.
\afterreview{
\subsubsection*{Ring Resonator}
}
The last structure is a micro-ring resonator. It is a racetrack shaped cavity whose length is $43 \: \mu$m and which is coupled to a narrow waveguide \afterreview{through evanescent wave coupling (section of both the ring and the waveguide are $500\times220 \: \textrm{nm}^2$)}. This narrow waveguide is linked to the input grating coupler by a 4.2~mm long broad (3 $\mu$m) waveguide. The cavity has been designed to obtain a free spectral range around 12~nm. After fabrication \afterreview{of our chips with different UV exposure doses}, a specific ring has been selected to get an agreement between the pump wavelength resonance and the bandwidth of our filtration lines (see next section). Linear properties of the cavity are presented in Fig. \ref{fig:Ring}.  There is a single resonance at $\lambda_{-1}=1528.0$nm, a split resonance at $\lambda_{0}=1540.0$ and $\lambda'_{0}=1540.1$~nm, and a split resonance at $\lambda_{1}=1552.1$ and $\lambda'_{1}=1552.2$~nm. The FWHM of the resonances (not taking into account the splitting) is 35~pm. \afterreview{Double resonances are a common property of microring resonator when they are coupled to a unique waveguide, it is a consequence of the symmetry breaking induced by the waveguide~\cite{kienmicrodisk}.} As the free spectral range is 12~nm, this give a finesse around $350$. We also estimate that on resonance the field enhancement factor described in the previous section $F_{p,s,i}$ is $F=110\pm 20$, where the uncertainty comes from the finite depth of the absorption peak in the transmission spectrum. With the pump at $\lambda_0$, photon pairs can be generated on the resonances at $\lambda_{-1}$ and $\lambda_1$ while ensuring energy conservation.

\section{Experiment}
\begin{figure}[!t]
\begin{center}
\includegraphics[width=12cm]{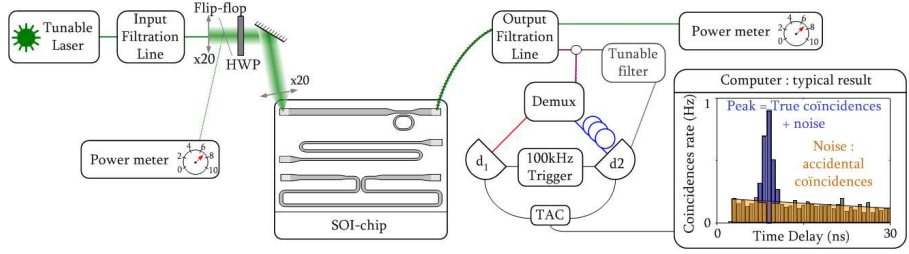}
\caption{Experimental Setup:
Tunable laser is made of an Agilent 81600B laser with picometer resolution followed by a homemade Erbium-Doped Fiber Amplifier.
\afterreview{Input Filtration is made of fiber bragg gratings (FBG), circulators, and 100GHz DWDM commercial filters.} It suppresses 150~dB outside of the pump band [1538.9-1540.6]~nm. 
HWP is an Half-Wave Plate.
Loss due to in and out coupling were $8\pm1$ ($7.5\pm1$)~dB for straight waveguide and Sagnac Interferometer (ring).
\afterreview{Output Filtration is made of 2 FBGs and two 200GHz DWDM commercial filters.} It suppresses 150~dB on the pump band; it induces $2.2$~dB of loss (to which one should add a 1~dB excess loss in a restricted band [1537-1534]~nm of the filtration line).
\afterreview{A first version of the setup} uses a demultiplexer  to separates Stokes band [1542-1558]~nm from anti-Stokes band [1523-1538]~nm. It is made of commercial CWDM filters and it induces 1~dB (2~dB) loss on Stokes (anti-Stokes) band. \afterreview{For the second version of the setup, the demultiplexer is made of DWDM commercial components and separates narrower Stokes band [1551.5-1552.1] from anti-Stokes band [1528.8-1528.35] with less than 1~dB of loss.}
d1 and d2 are commercial ID-quantique Avalanche Photodiodes (APD) with gate duration of $50$~ns and operationg at 100~kHz (d2 is trigged with a delay with respect to d1, this delay corresponds to the optical delay). \afterreview{For the first setup, detectors are ID-200 model with dark count rate of $5.6\ 10^{-5}$ and $4.4\ 10^{-5}$ per ns, and detection efficiency of 10$\%$. For the second setup, detectors are ID-201 model with detection efficiency set to 10$\%$ and 15$\%$ while dark counts are $1.4\ 10^{-5}$ and $3\ 10^{-5}$ per ns.}
TAC : time to amplitude converter. Time resolution of the coincidence detection system is $1.5$~ns.
Tunable filter has $6.5$~dB loss.
}
\label{fig:setup}
\end{center}
\end{figure}
Our experiment is based on a time coincidence measurement realized for various pump powers and for 3 different silicon nanostructures.
The setup of our experiment is depicted in Fig. \ref{fig:setup}.
\\
\indent
In the setup, the pump beam comes from an intense CW-laser.
For the straight waveguide and SLI experiments, the laser wavelength was taken to be 1539.6~nm, while for the ring resonator experiments, the pump wavelength was on the ring resonance at $1540\pm0.2$~nm (variation comes from resonnance shift due to pump power or external temperature change). The pump laser has to be filtered to suppress noise at Stokes and anti-Stokes frequencies. The pump beam is then coupled out of the all-fiber filtration line to minimize Raman scattering; pump beam's polarization is aligned on the TE-like mode of the silicon waveguide; and in/out coupling with the waveguide is ensured by grating couplers etched on both end of the nanostructure. \afterreview{Fiber-to-fiber transmission of the pump beam through the Si-waveguide is 20~dB.} The output filtration line suppresses \afterreview{150~dB of the} the pump beam. Afterward Stokes and anti-Stokes photons are spatially separated thanks to a demultiplexer, and sent into single photon detectors leading to detection coincidences.
\afterreview{Two different setups have been used which differ only by the detectors and the demultiplexer; first one has been used for straight waveguide and SLI characterization while the second has been used for ring characterisation and comparison with the straight waveguide.}
\\
\indent
Coincidences between Stokes and anti-Stokes detection are collected for hundreds of events, and the delay between the detections is recorded. Distinction between accidental coincidences and coincidences due to correlated photons is obtained by adding a systematic delay on the Stokes photon. The characteristic result of the experiment is a figure of coincidences versus delay between Stokes and anti-Stokes detection, see inset in Fig. \ref{fig:setup}. The peak corresponds to coincidences due to pairs (purple), while the background in other time-bins (brown) is the noise coming from destroyed photon pairs and dark counts of avalanche photodiodes. Furthermore, both input and output pump power are registered for each measurement to make sure the Fiber-to-Fiber loss does not change during the experiment. Stokes and anti-Stokes flux have also been measured for each experimental points. This give the flux of photons which may differ from the flux of photon pairs if there is another source of photons in the system. \afterreview{For the first setup, detection efficiency taking into account detector quantum efficiency, out coupling loss and filtration loss is $-21.2\pm1$~dB and $-22.2\pm1$dB for anti-Stokes and Stokes photons respectively. As the gate duration is 50~ns and the trigger rate 100~kHz, the detected coincidences are smaller than the generated pair flux by a factor of $2.5 \: 10^{-7}$. For second setup, detection efficiency reachs $-20.7$~dB and $-19$~dB for respectively Stokes and anti-Stokes photons, which results in a factor of  $5.4 \: 10^{-7}$ relating genereted pairs flux to detected coincidences rate.}
\\
\indent
The main features with which we characterize the photon pair sources are the photon pair flux and the Signal-to-Noise (SNR) ratio. The \textit{detected} photon pair flux is strongly dependent \afterreview{on efficiency and noise of detectors and also on losses. We therefore focussed on reconstructed \textit{generated} photon pair flux in the Si-structure as this is a more intrinsic characterization of the source than the \textit{detected} photon pair flux. Despite the Signal-to-Noise ratio  not characterizing the source in itself, it can be compared to predicted values that takes into account losses and dark counts.
Discrepancy with the predicted value will therefore be attributed to noise affecting the pair generation.} To make this comparison easier (see appendix) we have defined  the SNR as the number of events in the coincidence peak (all events combined in a single time bin of $0.5$~ns) divided by the number of events in this time-bin if the peak is erased, i.e.
\begin{equation}
\textrm{SNR} = \frac{c_a+c_p}{c_a}
\label{eq:SNR}
\end{equation}
where $c_a$ is the number of accidental coincidences in a noisy time-bin and $c_p$ the number of coincidences due to correlated photons in the corresponding time-bin.
\\
\indent
Finally, spectral properties of photon emission has also been investigated. We have observed the photon flux just after the output filtration line with an Optical Spectrum Analyzer or thanks to \afterreview{a unique APD and tunable filters with FWHW of 1.2~nm and 20~pm.}


\section{Results}
\subsection{Straight Waveguide}\label{sec:stwg}
\begin{figure}[!t]\begin{center}
\psfrag{x0}[cc][cc][1][0]{\footnotesize{0}}
\psfrag{x1}[cc][cc][1][0]{\footnotesize{10}}
\psfrag{x2}[cc][cc][1][0]{\footnotesize{20}}
\psfrag{x3}[cc][cc][1][0]{\footnotesize{30}}
\psfrag{x4}[cc][cc][1][0]{\footnotesize{40}}
\psfrag{y0}[cr][cr][1][0]{\footnotesize{0}}
\psfrag{y1}[cr][cr][1][0]{\footnotesize{4}}
\psfrag{y2}[cr][cr][1][0]{\footnotesize{8}}
\psfrag{y3}[cr][cr][1][0]{\footnotesize{12}}
\includegraphics[width=5.5cm]{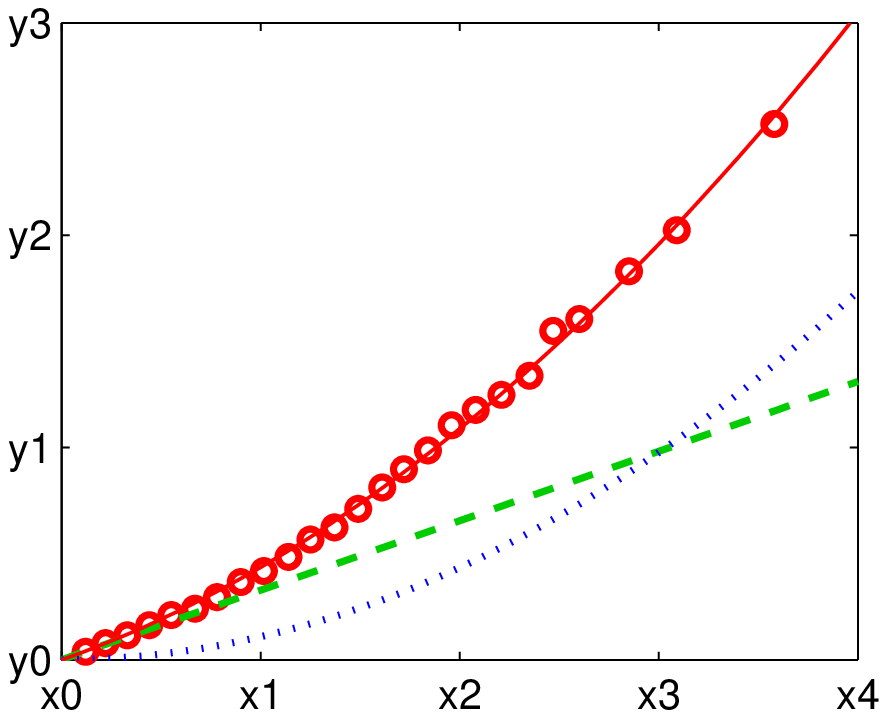}
\put(-122,-5){\footnotesize{Power before incoupling (mW)}}
\put(-158,22){\rotatebox{90}{\footnotesize{Stokes Photon Flux (kHz)}}}
\put(-30,100){\footnotesize{(a)}}
\hspace{0.1cm}
\psfrag{x0}[cc][cc][1][0]{\footnotesize{0}}
\psfrag{x1}[cc][cc][1][0]{\footnotesize{2}}
\psfrag{x2}[cc][cc][1][0]{\footnotesize{4}}
\psfrag{x3}[cc][cc][1][0]{\footnotesize{6}}
\psfrag{y0}[cr][cr][1][0]{\footnotesize{0}}
\psfrag{y1}[cr][cr][1][0]{\footnotesize{30}}
\psfrag{y2}[cr][cr][1][0]{\footnotesize{60}}
\psfrag{y3}[cr][cr][1][0]{\footnotesize{90}}
\includegraphics[width=5.5cm]{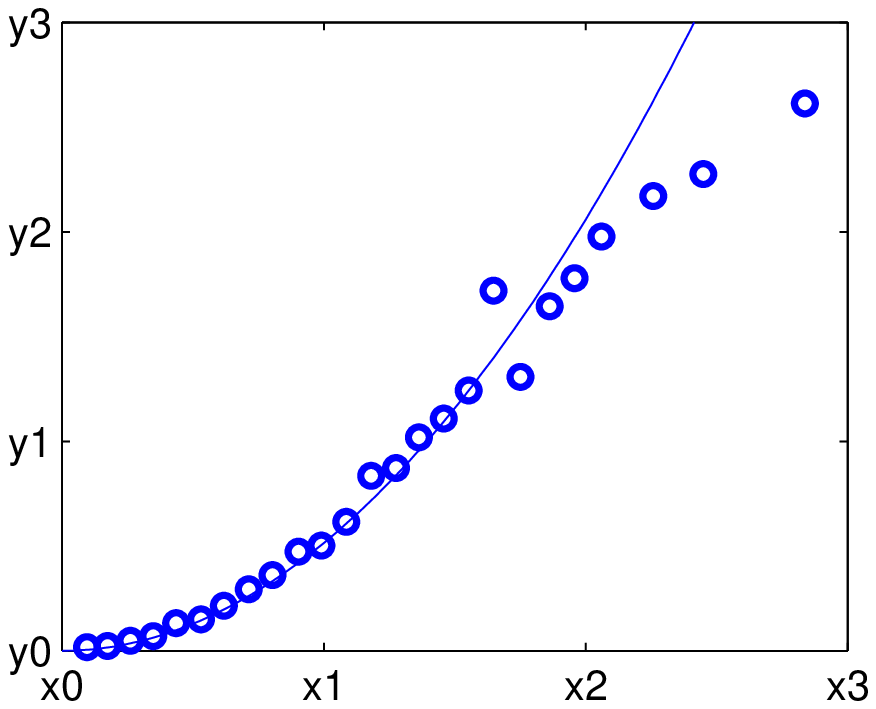}
\put(-118,-5){\footnotesize{Input Pump Power (mW)}}
\put(-158,19){\rotatebox{90}{\footnotesize{Generated Pair Flux (MHz)}}}
\put(-30,100){\footnotesize{(b)}}
\hspace{0.1cm}
\psfrag{x0}[cc][cc][1][0]{\footnotesize{0}}
\psfrag{x1}[cc][cc][1][0]{\footnotesize{30}}
\psfrag{x2}[cc][cc][1][0]{\footnotesize{60}}
\psfrag{x3}[cc][cc][1][0]{\footnotesize{90}}
\psfrag{y0}[cr][cr][1][0]{\footnotesize{0}}
\psfrag{y1}[cr][cr][1][0]{\footnotesize{3}}
\psfrag{y2}[cr][cr][1][0]{\footnotesize{6}}
\psfrag{y3}[cr][cr][1][0]{\footnotesize{9}}
\psfrag{y4}[cr][cr][1][0]{\footnotesize{12}}
\includegraphics[width = 5.5cm]{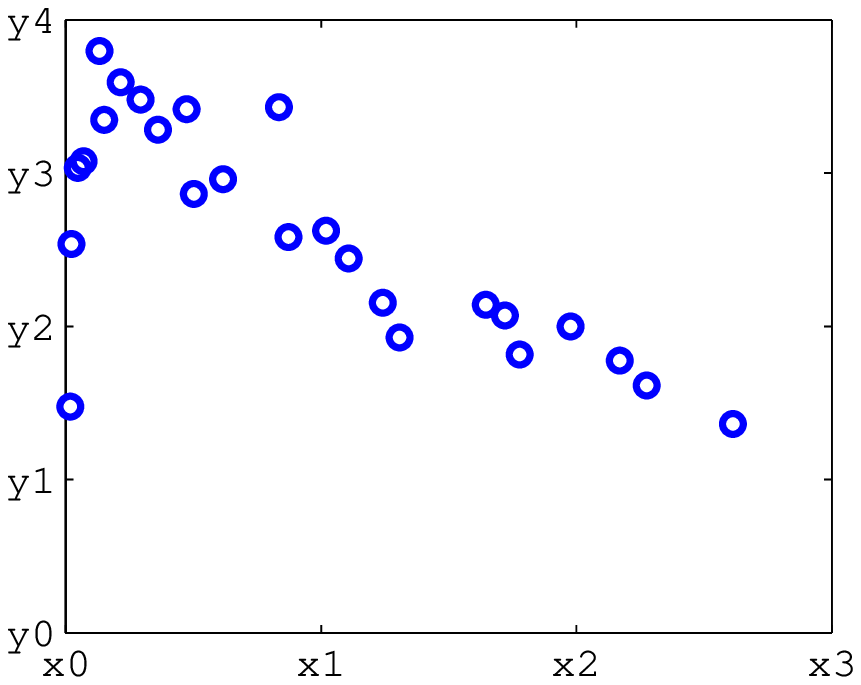}
\put(-120,-5){\footnotesize{Generated Pair Flux (MHz)}}
\put(-154,24){\rotatebox{90}{\footnotesize{Signal-to-Noise Ratio}}}
\put(-30,100){\footnotesize{(c)}}
\hspace{0.3cm}
\psfrag{x0}[ct][cc][1][0]{\footnotesize{1530}}
\psfrag{x1}[ct][cc][1][0]{\footnotesize{1535}}
\psfrag{x2}[ct][cc][1][0]{\footnotesize{1540}}
\psfrag{x3}[ct][cc][1][0]{\footnotesize{1545}}
\psfrag{x4}[ct][cc][1][0]{\footnotesize{1550}}
\psfrag{y0}[cr][cr][1][0]{\footnotesize{0}}
\psfrag{y1}[cr][cr][1][0]{\footnotesize{100}}
\psfrag{y2}[cr][cr][1][0]{\footnotesize{200}}
\psfrag{y3}[cr][cr][1][0]{\footnotesize{300}}
\psfrag{y4}[cr][cr][1][0]{\footnotesize{400}}
\psfrag{y5}[cr][cr][1][0]{\footnotesize{500}}
\includegraphics[width=5.5cm]{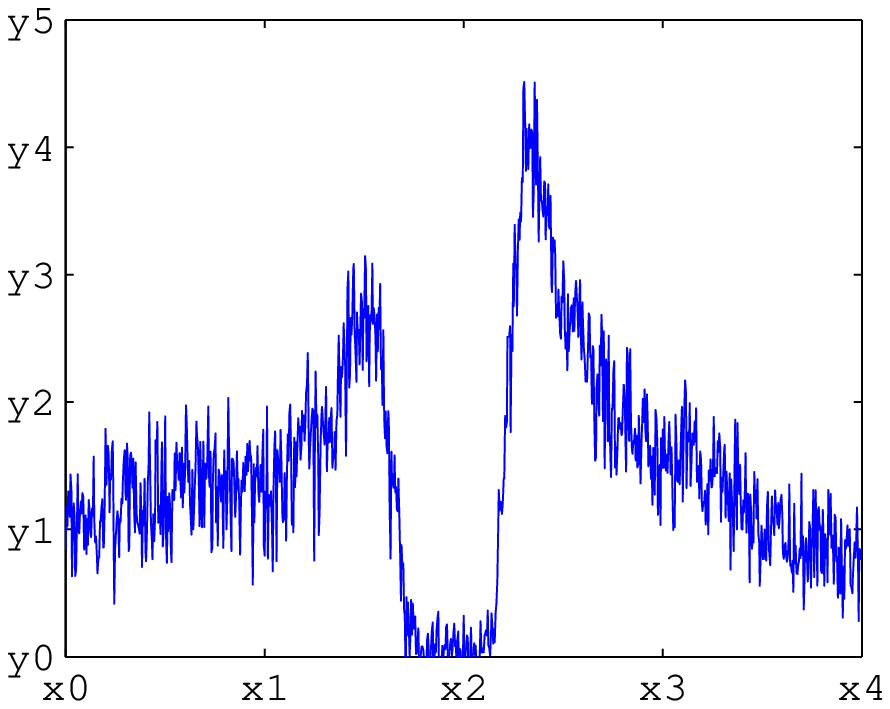}
\put(-105,-5){\footnotesize{Wavelength (nm)}}
\put(-160,17){\rotatebox{90}{\footnotesize{Emission spectrum (fW/nm)}}}
\put(-30,100){\footnotesize{(d)}}
\caption{Photon pair generation in straight waveguide. (a)~: Collected photon flux (dark counts are subtracted) in the Stokes band  versus the input pump power (\textcolor{red}{o}). Second order polynomial fit (\textcolor{red}{-}) and corresponding quadratic (\textcolor{blue}{$\cdot \cdot $}) and linear (\textcolor{green}{$--$}) contribution. The behavior is similar in the anti-Stokes band (figure not shown).
(b)~: Generated Photon pair flux (\afterreview{i.e. true coincidences rate corrected for loss, detection inefficiency, and time detection rate}) versus pump power in the waveguide. Experimental points (\textcolor{blue}{o}) are fitted (\textcolor{blue}{-}) by the flux given by Eq. (\ref{eq:flux}) with a bandwidth of 18~nm, an effective propagation distance of 4~mm and a correction factor of 1.4.
\afterreview{The factor relating generated pair flux and detected coincidence rate is $4 \: 10^{6}$}.
(c)~: Evolution of the SNR with respect to the generated pairs flux.
(d)~: Emission spectra measured with a conventional Optical Spectrum Analyzer (OSA) for a 10~mW pump power inside the waveguide. Noise from the OSA is subtracted and data are average over 10 scans. filtration line.
The asymmetry between Stokes and anti-Stokes frequencies is due to asymmetric losses in the output filtration line.
Note that panels (a), (b), (c) are taken with the first setup described in Fig. \ref{fig:setup}.
}
\label{fig:siW}
\end{center}\end{figure}
For our 11.3~mm straight waveguide the effective length is $L_{eff} \equiv L \exp{(-\alpha L)}=4 \: \textrm{mm}$. In such a waveguide, we estimated the flux from Eq. (\ref{eq:flux}) to be $f=3. \: 10^{-5} \textrm{ photon}/(\textrm{Hz.s})$ over a bandwidth wider than 40~nm for a 5~mw pump power, corresponding to a total photon pair production rate of 78~MHz.
The results are reported in  Fig. \ref{fig:siW} (for an early version of these results see~\cite{IEEELEOS}).
First the flux of photons is plotted as a function of pump power in Fig. \ref{fig:siW}(a). Experimental points are fitted by a second order polynomial curve which shows an unexpected significant linear contribution. Figure \ref{fig:siW}(b) shows the evolution of the generated pair flux versus the pump power in the waveguide.  The curve shows clearly a saturation at high pump power that correspond to detection flux close to 1 photon per detection gate. Experimental points are fitted by theoretical prediction Eq. (\ref{eq:flux}) that gives reasonable agreement. The observed Signal-to-Noise Ratio (SNR) is plotted in Fig. \ref{fig:siW}(c) as a function of the pump power. The maximum value obtained is SNR$=11.3$ at 9~MHz emission rate which is significantly less than the value of 69 expected for a 7~MHz as predicted by Eq. (\ref{eq:SNR}) (see appendix). This is partly due to linear scattering inside the waveguide (we estimate this around 2 or 3~dB). Nevertheless, the main factor that reduces the SNR is certainly related to uncorrelated photons emitted by the process responsible for the linear contribution to the photon flux. \afterreview{The SNR is higher in previously reported experiments mainly because of the higher collection and detection efficiency. In~\cite{kumar} and~\cite{ntt}, the discrepancy with theory has not been presented but seems to be in the same range (maybe for a different reason) as in our experiment.}
\\
\indent
The spectrum of emitted photons is plotted in Fig. \ref{fig:siW}(d). It shows an asymmetric peak with a dip that correspond to the block band response of the output filtration line. Asymmetry is due to asymmetric loss in the output filtration line. The non constant behavior of the emission (i.e. the peak) versus wavelength is unexpected and disagrees with theory Eq. (\ref{eq:flux}).

\subsubsection*{\textbf{Discrepancy with theoretical predictions}}
Our experimental results differ from our theoretical expectations in several respects.\\
First of all in the spectrum of Fig. \ref{fig:siW}(d) there is an unexpected peak near the pump wavelength, whereas theory Eq. (\ref{eq:flux}) predicts a flat spectrum over the investigated spectral range.   We believe the photon emission spectrum is the sum of a broadband flat spectrum due to photon pairs generation and a narrow band spectrum of unknown origin. \\
Second the flux of photons as reported in  Fig. \ref{fig:siW}(a) shows a clear linear component. Additional measurements indicate that 10 percents of this linear contribution is due to Raman scattering in both input and output filtration line made of fiber components. The remaining 90$\%$ are unexplained.\\
These two unexpected features are probably related, and might be due to a spontaneous emission similar to Raman scattering. For instance Raman scattering might be generated in the silica layer underneath the silicon ribbon. Others explanations could be spontaneous process involving carriers such as Free-Carrier-induced dispersion or forward Brillouin spontaneous scattering. Further investigation in a nanosecond pulsed regime could invalidate or confirm the possible role of carriers. Note that neither the linear contribution to the emitted flux nor the photon emission spectrum has been exhibited in previously reported experiments.
\subsection{Sagnac Loop Interferometer}
\begin{figure}[!ht]\begin{center}
\psfrag{x0}[cl][cc][1][0]{\footnotesize{0}}
\psfrag{x1}[cc][cc][1][0]{\footnotesize{2.5}}
\psfrag{x2}[cc][cc][1][0]{\footnotesize{5}}
\psfrag{x3}[cc][cc][1][0]{\footnotesize{7.5}}
\psfrag{x4}[cc][cc][1][0]{\footnotesize{10}}
\psfrag{y0}[cr][cr][1][0]{\footnotesize{0}}
\psfrag{y1}[cr][cr][1][0]{\footnotesize{1}}
\psfrag{y2}[cr][cr][1][0]{\footnotesize{2}}
\psfrag{y3}[cr][cr][1][0]{\footnotesize{3}}
\psfrag{y4}[cr][cr][1][0]{\footnotesize{4}}
\psfrag{y5}[cr][cr][1][0]{\footnotesize{5}}
\includegraphics[width=5.5cm]{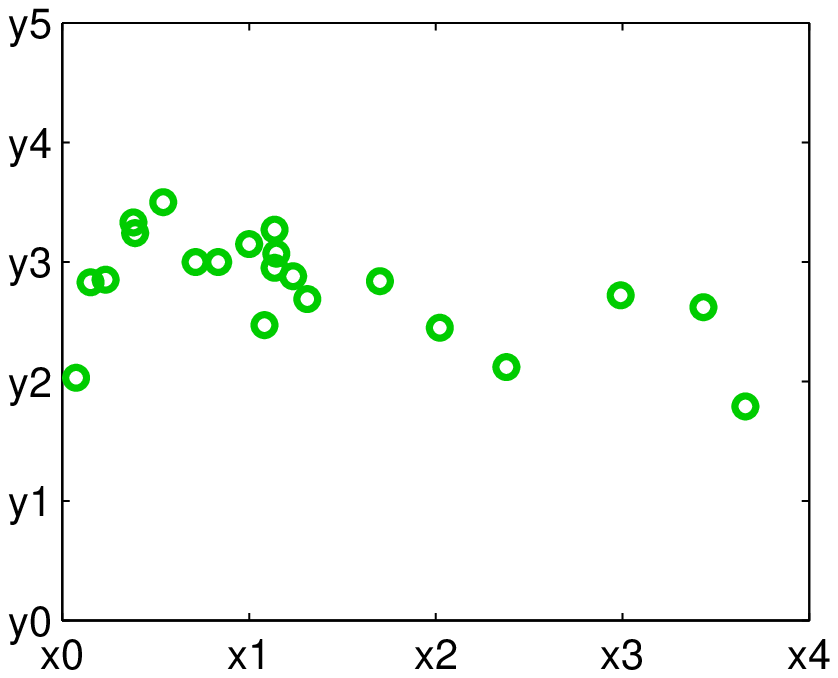}
\put(-120,-5){\footnotesize{Generated Pair Flux (MHz)}}
\put(-154,27){\rotatebox{90}{\footnotesize{Signal-to-Noise Ratio}}}
\put(-30,100){\footnotesize{(a)}}
\hspace{0.3cm}
\psfrag{x0}[cl][cc][1][0]{\footnotesize{0.1}}
\psfrag{x1}[cc][cc][1][0]{\footnotesize{1}}
\psfrag{x2}[cc][cc][1][0]{\footnotesize{10}}
\psfrag{x3}[cc][cc][1][0]{\footnotesize{100}}
\psfrag{y0}[cr][cr][1][0]{\footnotesize{0.1}}
\psfrag{y1}[cr][cr][1][0]{\footnotesize{1}}
\psfrag{y2}[cr][cr][1][0]{\footnotesize{10}}
\psfrag{y3}[cr][cr][1][0]{\footnotesize{100}}
\psfrag{y4}[cr][cr][1][0]{\footnotesize{1000}}
\includegraphics[width=5.5cm]{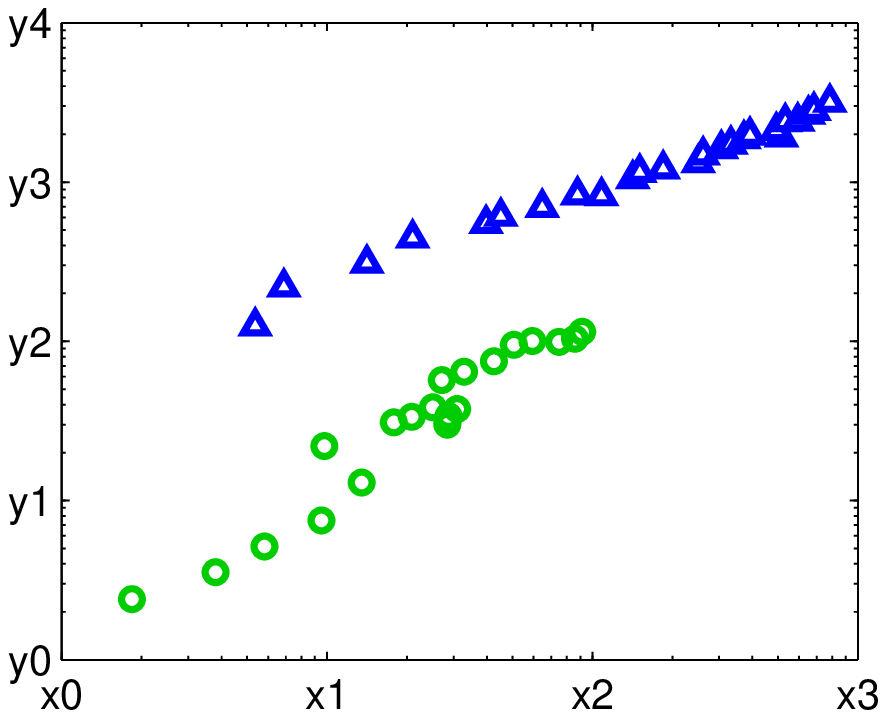}
\put(-113,-5){\footnotesize{Generated Pair flux (MHz)}}
\put(-159,17){\rotatebox{90}{\footnotesize{Output Pump Power ($\mu$W)}}}
\put(-30,100){\footnotesize{(b)}}
\caption{Photon pair generation in a Sagnac loop interferometer.
(a)~: Signal-to-Noise Ratio versus the generated pair flux inside the Sagnac loop. SNR is lower than in the straight waveguide. This is mainly due to the lower efficiency of the photon pair generation.
(b)~: Output pump power (\afterreview{measured after the first filter of the output filtration line}) versus generated photon pair flux from the $11.3$~mm long straight waveguide (\textcolor{blue}{$\triangle$}) and from the Sagnac loop interferometer (\textcolor{green}{o}). The output pump power for a given pair flux is 6 to 16~dB lower for the Sagnac loop than for the straight waveguide despite the lower efficiency of photon pair generation in the Sagnac loop.
}
\label{fig:Sagnac}
\end{center}\end{figure}
Results obtained for the Sagnac loop interferometer are presented in Fig. \ref{fig:Sagnac}. Left panel (a) shows the SNR versus the generated pair flux : it is 4 times lower than in the straight waveguide. This effect is mainly due to the lower efficiency of the photon pair generation which makes the noise contribution higher. We measured that generation efficiency is 7 times lower for the same pump power (which can be partly explained by the fact that the pump power is split in two at the coupler). Furthermore, some pairs have probably been produced inside the input broad waveguide : those pairs mainly contribute to noise because they experience higher losses. Notwithstanding the above  issues, the Sagnac interferometer reduced the pump power by 6 to 16~dB for given photon pair flux, see Fig. \ref{fig:Sagnac}(b). This last result represents an interesting improvement as it allows in principle to add other components (interferometers for instance) after the pair source while localizing the  pair generation.
The variation of pump extinction with pair flux is partially due to a slight nonlinear behavior of pump extinction versus input pump power.
\subsection{Ring Resonator}
\begin{figure}[!t]\begin{center}
\psfrag{x0}[ct][cc][1][0]{\footnotesize{1530}}
\psfrag{x1}[ct][cc][1][0]{\footnotesize{1540}}
\psfrag{x2}[ct][cc][1][0]{\footnotesize{1550}}
\psfrag{y0}[cr][cr][1][0]{\footnotesize{-30}}
\psfrag{y1}[cr][cr][1][0]{\footnotesize{-20}}
\psfrag{y2}[cr][cr][1][0]{\footnotesize{-10}}
\psfrag{y3}[cr][cr][1][0]{\footnotesize{0}}
\psfrag{w0}[cr][cr][1][0]{\tiny{0}}
\psfrag{w1}[cr][cr][1][0]{\tiny{1}}
\psfrag{z0}[ct][cc][1][0]{\tiny{1527.8}}
\psfrag{z1}[ct][cc][1][0]{\tiny{1528.2}}
\psfrag{z2}[ct][cc][1][0]{\tiny{1539.8}}
\psfrag{z3}[ct][cc][1][0]{\tiny{1540.2}}
\psfrag{z4}[ct][cc][1][0]{\tiny{1551.9}}
\psfrag{z5}[ct][cc][1][0]{\tiny{1552.3}}
\includegraphics[width=5.5cm]{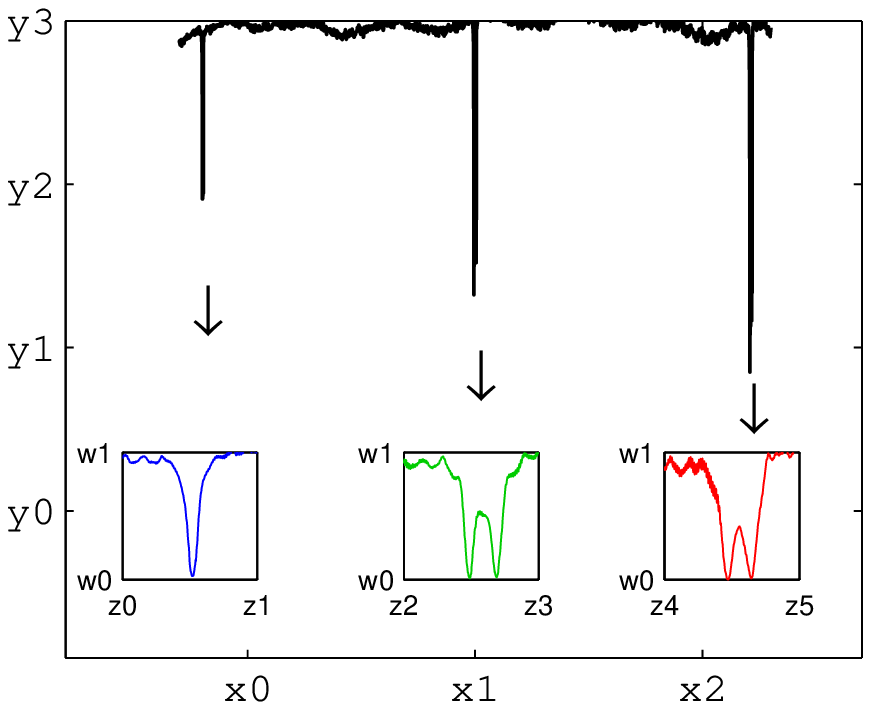}
\put(-105,-5){\footnotesize{Wavelength (nm)}}
\put(-158,30){\rotatebox{90}{\footnotesize{Transmission (dB)}}}
\put(-30,100){\footnotesize{(a)}}
\hspace{0.3cm}
\psfrag{x0}[ct][cc][1][0]{\footnotesize{1520}}
\psfrag{x1}[ct][cc][1][0]{\footnotesize{1530}}
\psfrag{x2}[ct][cc][1][0]{\footnotesize{1540}}
\psfrag{x3}[ct][cc][1][0]{\footnotesize{1550}}
\psfrag{x4}[ct][cc][1][0]{\footnotesize{1560}}
\psfrag{y0}[cr][cr][1][0]{\footnotesize{0}}
\psfrag{y1}[cr][cr][1][0]{\footnotesize{0.25}}
\psfrag{y2}[cr][cr][1][0]{\footnotesize{0.5}}
\psfrag{y3}[cr][cr][1][0]{\footnotesize{0.75}}
\psfrag{y4}[cr][cr][1][0]{\footnotesize{1}}
\psfrag{z0}[tc][cc][1][0]{\footnotesize{1552}}
\psfrag{z1}[tc][cc][1][0]{\footnotesize{1552.3}}
\psfrag{w0}[br][cr][1][0]{\footnotesize{0}}
\psfrag{w1}[tr][cr][1][0]{\footnotesize{1}}
\includegraphics[width=5.5cm]{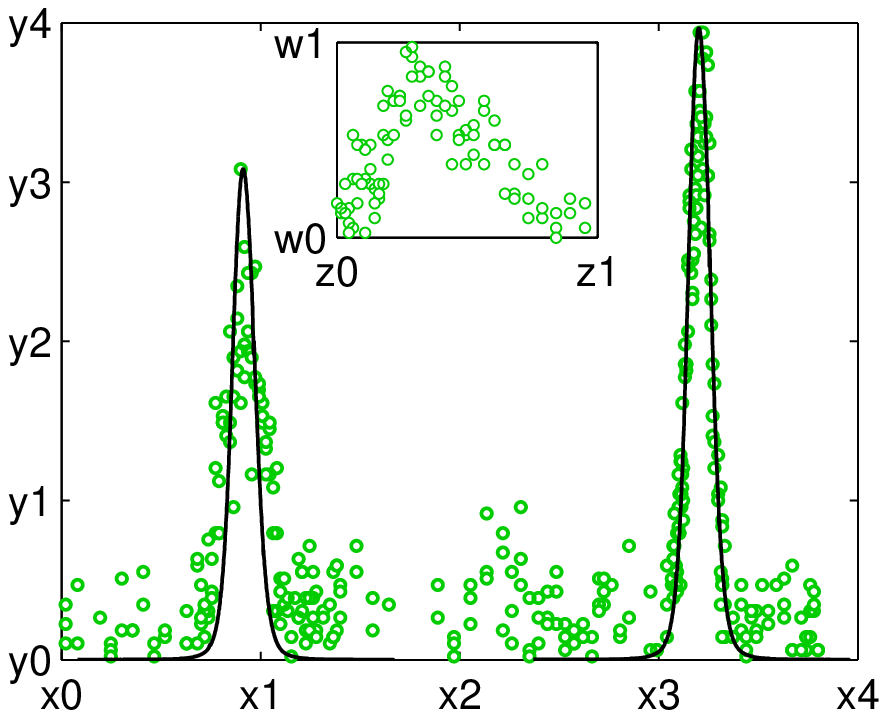} 
\put(-105,-8){\footnotesize{Wavelength (nm)}}
\put(-162,12){\rotatebox{90}{\footnotesize{Normalized emission spectrum}}}
\put(-30,100){\footnotesize{(b)}}
\hspace{0.1cm}
\psfrag{x0}[cl][cc][1][0]{\footnotesize{0.1}}
\psfrag{x1}[cc][cc][1][0]{\footnotesize{1}}
\psfrag{x2}[cc][cc][1][0]{\footnotesize{10}}
\psfrag{y0}[cr][cr][1][0]{\footnotesize{0.1}}
\psfrag{y1}[cr][cr][1][0]{\footnotesize{0.3}}
\psfrag{y2}[cr][cr][1][0]{\footnotesize{1}}
\includegraphics[width=5.5cm]{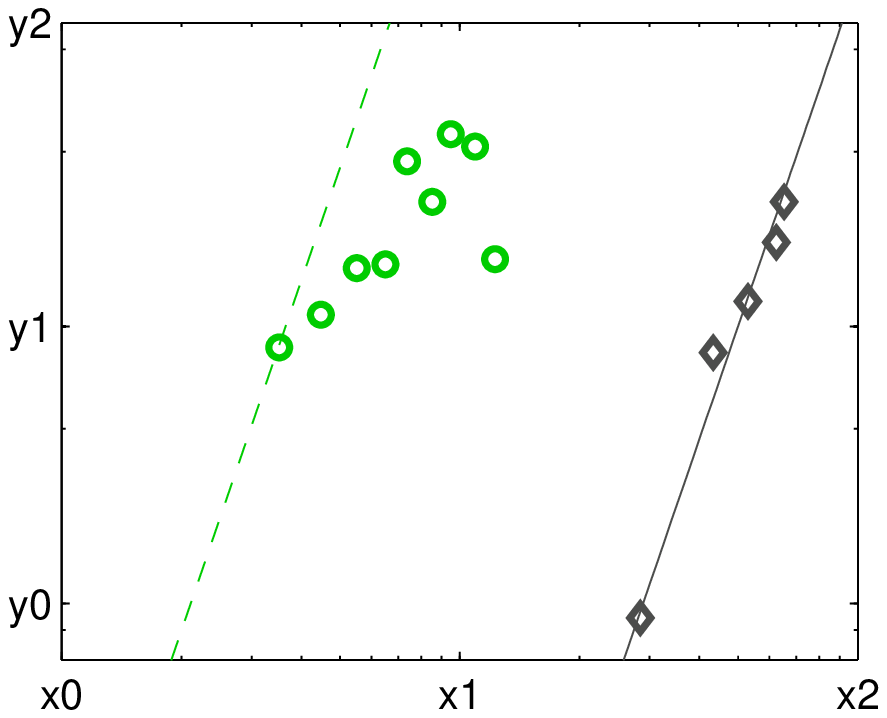}
\put(-110,-1){\footnotesize{Input Pump Power (mW)}}
\put(-158,18){\rotatebox{90}{\footnotesize{Generated Pair Flux (MHz)}}}
\put(-30,100){\footnotesize{(c)}}
\hspace{0.3cm}
\psfrag{x0}[cl][cc][1][0]{\footnotesize{0}}
\psfrag{x1}[cc][cc][1][0]{\footnotesize{0.25}}
\psfrag{x2}[cc][cc][1][0]{\footnotesize{0.5}}
\psfrag{x3}[cc][cc][1][0]{\footnotesize{0.75}}
\psfrag{y0}[cr][cr][1][0]{\footnotesize{0}}
\psfrag{y1}[cr][cr][1][0]{\footnotesize{10}}
\psfrag{y2}[cr][cr][1][0]{\footnotesize{20}}
\psfrag{y3}[cr][cr][1][0]{\footnotesize{30}}
\psfrag{y4}[cr][cr][1][0]{\footnotesize{40}}
\includegraphics[width=5.5cm]{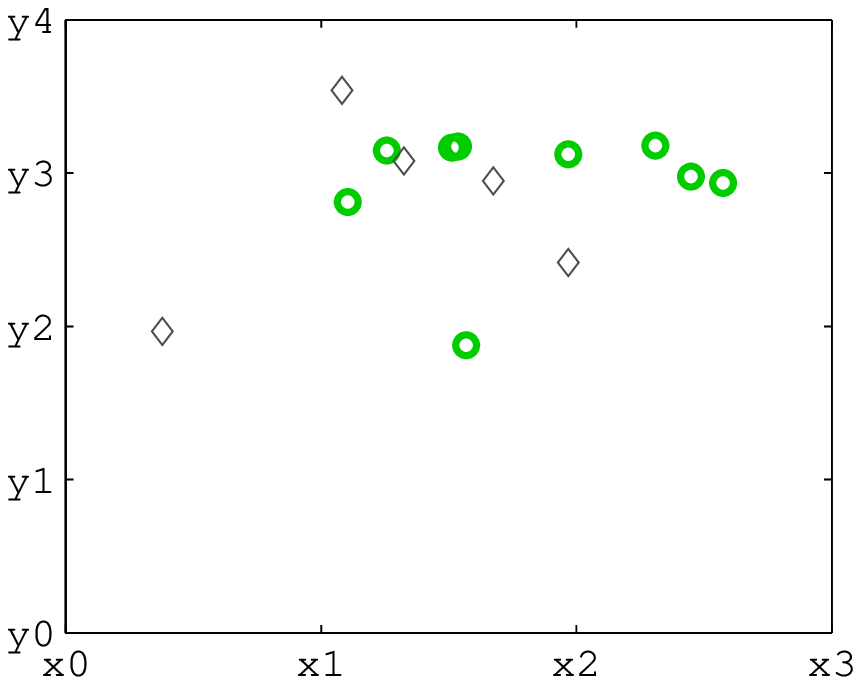}
\put(-120,-1){\footnotesize{Generated Pair Flux (MHz)}}
\put(-158,28){\rotatebox{90}{\footnotesize{Signal-to-Noise Ratio}}}
\put(-30,100){\footnotesize{(d)}}
\caption{Photon pair generation in micro-ring resonator. \afterreview{Note that  results presented in panels (a) and (b) differs from those in panels (c) and (d) by a temperature change due to external conditions. Note that panels (c) and (d) were taken with the second setup described in Fig. \ref{fig:setup}.}
(a)~: Transmission spectrum of the ring cavity. Insets: zoom on the resonances.
(b)~: Emission spectrum from the ring cavity. The measurement is made thanks to a tunable filter and a single photon detector for an estimated pump power at the input of the cavity of $0.4$~mW. The width of the peaks are limited by the linewidth of the tunable filter (continuous curves \afterreview{- FWHM $= 1.3$~nm}). The linewidth of the Stokes resonance has been investigated (inset) thanks to a very narrow tunable filter (FWHM = 20~pm) which shows the emission linewidth to be approximately 150~pm.
(c)~: Generated Photon Pair flux (i.e. coincidences only as in Fig. \ref{fig:siW}(b)) versus pump power for pumping on resonance (\textcolor{green}{o}, $\lambda_{Pump} \approx 1539.9\pm0.05$~nm). Quadratic curve (\textcolor{green}{$--$}) show that evolution of pair generation inside the ring resonator does not follow any clear law, see discussion in the main text. \afterreview{Grey diamonds (\textcolor{gris}{$\diamond$}) show the pair flux generated in the 11.3~mm long straight waveguide while filtered by the demultiplexer used for the ring. The corresponding quadratic fit is plotted in grey (\textcolor{gris}{-}).}
(d)~: SNR versus generated pair flux generated in the ring resonator (\textcolor{green}{$o$}) and in the 11.3~mm long straight waveguide  (\textcolor{gris}{$\diamond$}).
}
\label{fig:Ring}
\end{center}
\end{figure}
When the pump, signal and idler are all on resonance, Eq. \ref{eq:fluxring} predicts that the flux spectral density $f$  should be approximately 3700 times larger for the ring than in the straight waveguide of length $11.3$~mm (for identical pump power). The pair generation rate, when integrated over the width of the resonance of the signal and idler (the pump being on resonance), should be 11~MHz for a pump power of $0.4$~mW.  However for a pump power of $0.4$~mW, the power inside the ring should be $44$~mW, whereupon free carrier absorption becomes important, see Fig. \ref{fig:loss}, which should temper these conclusions.
\\
\indent
The experimental results are plotted in Fig. \ref{fig:Ring}.
The emission spectrum in Fig. \ref{fig:Ring}(b) suggests that the pairs are generated over a 150~pm bandwidth, rather than the 35~pm FWHM of the resonances measured in Fig. \ref{fig:Ring}(a). \afterreview{This is probably due to nonlinear absorption (see below).}
\\
\indent
The evolution of the generated photon pair flux as a function of pump power
is compared with the \afterreview{flux generated in the 11.3~mm long waveguide in Fig. \ref{fig:Ring}(c).
It shows that the required input pump power for a given flux is reduced by roughtly 10 dB with respect to the straight waveguide, which means a process efficiency 2 orders of magnitude higher.}
Note that for a pump power $P=0.4$~mW the pair generation rate \afterreview{is $0.3$~MHz}. Below we discuss the discrepancy with the theoretical prediction of $11$~MHz.
\\
\indent
From Figs. \ref{fig:Ring} (b) and Fig. \ref{fig:Ring}(c) we deduce that for the ring resonator the photon pair spectral density is \afterreview{2~MHz/nm} for a pump power of $400\: \mu$W.
From Fig. \ref{fig:Ring}(c) we deduce that this value is
more than 2 orders of magnitude higher than the expected value for the $11.3$~mm long straight waveguide at the same pump power.
\\
\indent
The SNR is presented in Fig. \ref{fig:Ring}(d) as a function of the generated pair flux. Taking into account collection efficiency and dark counts, the expected peak of the SNR curve should be 250 for the ring (225 for the straight waveguide), compared to the measured value of 30.

\subsubsection*{\textbf{Discussion}}
The observed photon pair flux in the micro-ring is significantly lower than predicted theoretically. The most natural interpretation is that, due to the large field enhancement, nonlinear losses in the ring are important. This broadens the resonances (as observed) and  decreases the field enhancement factors $F_{p,s,i}$. As these appear to the $4^{\textrm{th}}$ power in Eq. \ref{eq:fluxring} this can easily explain the discrepancy between the observed pair flux of $0.3$~MHz and the predicted flux of 11~MHz.
\\
\indent
Our experiment is probably also affected by shifts of the resonance due to free-carrier dispersion and thermal effects~\cite{bistab:AlmeidaLipson}.
Indeed, although for each point in Fig. \ref{fig:Ring}  we adjusted the pump wavelength to maximize the total flux, from one measurement to another we observed the pair flux to change by up to 2 dB. This can explain the scatter of the data in Figs. \ref{fig:Ring} (c) and \ref{fig:Ring} (d).
\\
\indent
Finally we attribute the lower than expected SNR in Fig. \ref{fig:Ring} (d) to the same source of noise that is discussed in the context of the straight waveguide (see Section \ref{sec:stwg} and Fig. \ref{fig:siW}).

\section{Conclusion}
The use of a continuous pump could make photon pairs based on SOI straight waveguides or  micro-ring resonators an attractive source for long distance quantum communication at telecommunication wavelengths, as this is much simpler and cheaper than a pulsed laser.
The resulting source could either be broadband (based on a straight waveguide) or narrow band (based on ring resonators). This application would however require that the noise sources uncovered in the present work be reduced.
\\
\indent
Note that in the future the laser could even be integrated on the chip itself, following the recent demonstrations of lasers bonded on Si chips~\cite{laseronchip}. If necessary the CW beam could also be modulated on chip, as in~\cite{MZmodulator}. Furthermore the nonlinear absorption could be reduced thereby allowing higher pump powers by using a PIN junction as in~\cite{RamanLaser}, and the coupling losses to optical fibers could be reduced to $\simeq 1$dB by using inverted nanotapers.
\\
\indent
The present work also suggests that SOI could be a promising platform on which to realize linear optics quantum computing~\cite{klm}. In the past the main limitation for working at telecommunication wavelengths was the low quality of the single photon detectors, but this has become much less of an issue with recent advances.
The use a Sagnac interferometer or a micro-ring resonator ensures that the photon pair generation is localized in a specific part of the chip. The source could then be combined on chip with wavelength demultiplexers, couplers and interferometers to realize the LOQC. In this way one could hope to parallel the recent advances on integrated LOQC realized using silica waveguides~\cite{quantumcircuits}.
\section*{Appendix: Signal-to-Noise Ratio}
We defined the Signal-to-Noise ratio as the number of events in the coincidence peak (all events combined in a single time bin of $0.5$~ns) divided by the number of events in this time-bin if the peak is erased. This definition allows us a comparison with theory that doesn't need to take into account lack of temporal resolution in our detection system. It is equivalent to :
\begin{equation}
\textrm{SNR} = \frac{c_a + c_p}{c_a}
\label{eq:SNR}
\end{equation}
with $c_a$ the number of accidental coincidences and $c_p$ the number of coincidences due to correlated pairs. If the photon pair flux is low enough (much less than one photon pair for each detection gate), one can easily evaluate $c_a$ and $c_p$ depending on dark counts $dk_{s,a}$ in each detector, total detection efficiency $\eta_{s,a}$ of Stokes and anti-Stokes photons, time-bin duration $\tau_b$, and pair generation rate $\gamma_e$. The number of accidental coincidences is given by the temporal integration over one time-bin duration of the product of detection probability (per unit of time) in the Stokes band times the detection probability (per unit of time) in the anti-Stokes band when those probability are uncorrelated.
The detection probability per unit of time depends on the dark count per unit of time $dk$, the flux of emitted photons $\gamma_e \tau_{b}$ and the detection efficiency $\eta$ (including losses). Thus detection probability per unit of time is expressed as :
\begin{equation}
p_{s,a}=\gamma_e  \eta_{s,a} + dk_{s,a} \qquad \textrm{($s,a$ stand for Stokes or anti-Stokes)}
\label{eq:p}
\end{equation}
As this probability remains constant over time, we can express the number of accidental coincidences during a time-bin $\tau_b$ as
\begin{equation}
c_a = (\gamma_e  \eta_s + dk_s) (\gamma_e  \eta_a + dk_a)  \tau_b^2
\label{eq:ca}
\end{equation}
The number of coincidences due to pairs is deduced in the same way, but taking into account that the Stokes and anti-Stokes photons
are produced simultaneously, hence the probability of detection of the anti-Stokes photon when the corresponding Stokes has been detected is  given by $\eta_a$. This yields:
\begin{equation}
c_p = \gamma_e  \eta_{s} \eta_a \tau_b
\label{eq:cp}
\end{equation}

\section*{Acknowledgments}
We acknowledge the support of the \textit{Fonds pour la formation \`a la Recherche dans l'Industrie et dans l'Agriculture} (FRIA, Belgium), of the \textit{Fond National pour la Recherche Scientifique}, of the Research Foundation-Flanders (FWO), of the \textit{Interuniversity Attraction Poles Programme} (Belgian Science Policy) under project Photonics@be IAP6-10, of the EU project \textit{QAP} contract 015848, of the \textit{Programme International de Coop\'eration Scientifique} PICS-3742, of the \textit{Groupement de Recherche Photonique Nonlin\'eaire et Milieu Microstructur\'es} GDR-3073, of the CNRS, and the \textit{conseil régional de Franche Comté}. We also thank Freddy Clavie and Lory Marchal for technical support.

\end{document}